\documentstyle[preprint,epsfig,aps,amssymb]{revtex}
\begin{document}
\draft
\tightenlines

\def\br{{\bf r}}
\def\bk{{\bf k}}
\def\bu{{\bf u}}
\def\bw{{\bf w}}
\def\brt{\br,t}
\def\bbrt{(\brt)}
\def\bprt{\bp,\brt}
\def\cphio{\Phi_0}
\def\beq{\begin{equation}}
\def\eeq{\end{equation}}
\def\bea{\begin{eqnarray}}
\def\eea{\end{eqnarray}}
\def\bna{\bbox{\nabla}}
\def\bp{{\bf p}}
\def\bv{{\bf v}}
\def\tn{\tilde n}
\def\tp{\tilde p}
\def\be{\bbox{\eta}}
\def\la{\langle}
\def\ra{\rangle}

\title{Transverse modes of a cigar-shaped Bose-Einstein condensate}
\author{B. Jackson and E. Zaremba}
\address{Department of Physics, Queen's University, Kingston, Ontario 
 K7L 3N6, Canada \\
brian@sparky.phy.queensu.ca}
\date{\today}
\maketitle
\begin{abstract}

We discuss the collective modes in a harmonically trapped, 
highly-elongated Bose condensed gas. The transverse breathing mode
exhibits a number of interesting features, such as the insensitivity of
the condensate mode frequency to the interaction strength, and the 
closeness of the frequency to that of the non-condensed thermal cloud in
the collisionless limit. Using finite temperature simulations, we show 
that these features are responsible for the very small damping rate
observed experimentally. Our results for the temperature dependence of
the damping rate and frequency shift are in excellent agreement with
experiment. We also demonstrate that the unusually small damping rate
does not arise for the $m=2$ mode or for more isotropic trap potentials,
suggesting further possible experimental tests of our theory. 
    
\end{abstract}

\section{Introduction}

The study of the finite temperature properties of trapped Bose-Einstein 
condensed gases touches on fundamental issues such as condensate growth
and relaxation processes in an isolated quantum degenerate system. 
As such, they have been the subject of ongoing research since soon
after the Nobel prize-winning discoveries of 1995. For example, early
experiments at JILA \cite{jin97} and MIT \cite{stamper-kurn98} measured 
the frequency and 
damping rates of collective excitations as a function of temperature. 
More recent experiments have studied the finite temperature behavior of
the scissors mode \cite{marago01}, and the radial breathing mode in a 
cigar-shaped condensate \cite{chevy02}. These experiments have in turn
stimulated a considerable amount of theoretical work.

Finite temperature theories have taken various forms. Early work 
focussed on solving Hartree-Fock-Bogoliubov (HFB) equations to find 
the excitation spectrum of the system 
\cite{hutchinson97,dodd98,hutchinson98}. These excited states, when 
thermally occupied, determine the equilibrium density distribution of 
the non-condensed thermal cloud. Unfortunately, this method is 
unsatisfactory when considering the low energy collective modes 
since it treats 
the thermal cloud as static. In reality the condensate oscillations 
will induce fluctuations in the thermal component, which then act back 
on the condensate leading to damping and associated
frequency shifts. One possible way of addressing this problem is to
treat the thermal cloud dynamics perturbatively. This method has been
pursued by a number of authors 
\cite{morgan00,rusch00,giorgini00,guilleumas00,reidl00}, and  
although good agreement has been found with some of the experimental 
results\cite{jin97,stamper-kurn98}, it is still unclear whether the
method is useful in situations where the full dynamics of the thermal 
cloud is important.

An alternative approach is to formulate a quantum kinetic theory, which
in principle treats the dynamics of the system fully and consistently. 
This includes second-order collisional processes, which lead to 
equilibration of the system. Kinetic theories for dilute Bose gases have 
been formulated by Stoof \cite{stoof99}, Gardiner and co-workers 
\cite{gardiner00}, 
Walser {\it et al.}, \cite{walser99} and Zaremba, Nikuni and Griffin (ZNG) 
\cite{zaremba99}. Finally, there is the possibility
of treating the evolution of the system using Fokker-Planck equations for 
quasiprobability distributions. A specific example of this is the truncated
Wigner representation, or classical field method \cite{sinatra01,davis01},
which is valid when the occupation of the relevant states is much greater
than unity. 

In this paper we shall present results of solving the ZNG equations 
using a method detailed in a previous publication \cite{jackson02b}. 
Earlier applications of the method yielded results in good agreement 
with the scissors mode experiment at Oxford \cite{marago01,jackson01} 
as well as with the quadrupole modes in the JILA 
experiment~\cite{jin97,jackson02a}. Here we focus instead on
the ENS experiment~\cite{chevy02}, adding more details to the discussion
of our simulations in Ref.~\cite{jackson02c}. The ENS experiment
studied the radial breathing mode of an elongated condensate, and found
unusually small damping rates and frequency shifts as compared to other 
experiments. We explain this behaviour in terms of an accidental 
degeneracy between the condensate and thermal cloud oscillation 
frequencies. In Section II we review the ZNG theory and our numerical 
scheme, while in Section III we discuss the collective modes in an 
elongated harmonic trap. In Section IV we present results from our 
finite temperature simulations for the tranverse breathing and 
quadrupole modes in this geometry. For
the latter, the condensate and thermal cloud frequencies are no 
longer degenerate and we find a much larger damping rate and frequency
shift. We also show that larger damping rates arise when the degeneracy
is lifted in more isotropic trap potentials.
Finally, Section V presents our conclusions. 

\section{ZNG Theory}

We first give an overview of the ZNG theory which forms the basis of the
numerical simulations described in this paper. Further details of the 
formalism, including the approximations made, can be found in Ref.\ 
\cite{zaremba99}. In the theory two coupled equations are used to 
describe the condensate and thermal cloud dynamics. In particular, the 
condensate wavefunction $\Phi (\brt)$ evolves according to 
the generalized Gross-Pitaevskii (GP) equation
\beq
  i\hbar \frac{\partial}{\partial t} \Phi(\brt) = \left (
 -\frac{\hbar^2 \nabla^2}{2m} + U_{\rm ext} (\brt) + g[n_c (\brt)+2\tilde{n}
 (\brt)] - iR(\brt) \right) \Phi(\brt).
\label{eq:GP-gen}
\eeq
The condensate interacts locally with other condensate atoms and with
the non-condensate, with interactions parametrized by 
$g=4\pi\hbar^2 a/m$, where $a$ is the $s$-wave scattering length and $m$ 
is the mass of the atoms. These mean-field interactions involve the
condensate density $n_c (\brt) = |\Phi(\brt)|^2$ and the non-condensate
density $\tilde n(\brt)$.
The term $U_{\rm ext} (\brt) = m[\omega_{\bot}^2 (x^2 +y^2)
+\omega_z^2 z^2]/2$ represents the trap potential. The frequencies
in the radial ($\omega_{\bot}$) and axial ($\omega_z$) directions are 
generally different, and the ratio $\lambda=\omega_z/\omega_{\bot}$ 
conveniently parameterizes the trap anisotropy.

The thermal cloud is described in terms of a semiclassical Wigner
distribution $f(\bprt)$ which defines the non-condensate density
according to 
\beq
 \tilde{n} (\brt) = \int \frac{{\rm d}\bp}{(2\pi \hbar)^3} f(\bprt)\,.
\label{eq:normdens}
\eeq 
This distribution function satisfies the kinetic equation
\begin{equation}
 \frac{\partial}{\partial t} f (\bprt) + \frac{\bp}{m} \cdot \nabla
 f (\bprt) - \nabla U(\brt) \cdot \nabla_{\bp} f (\bprt)
 = C_{12}[f] + C_{22} [f]\,,
\label{eq:kinetic}
\eeq
where the term $U(\br,t)=U_{\rm ext} (\br)+ 2g[n_c (\br,t)+\tilde{n} 
(\br,t)]$ reflects the combined effect of the trap and mean-field 
potentials on the thermal cloud evolution. Interactions also enter in 
the terms on the 
right-hand side, which are of order $g^2$ and correspond to binary
collision events between thermal atoms ($C_{22}$) and between thermal 
and condensate atoms ($C_{12}$). The second process is particulary important 
in the context of condensate growth, since it leads to a transfer of atoms 
between the condensate and thermal cloud. It also accounts for the non-Hermitian
term in the GP equation (\ref{eq:GP-gen}) which is given by
\beq
 R (\brt) = \frac{\hbar}{2n_c} \int \frac{{\rm d}\bp}{(2\pi\hbar)^3}
 C_{12} [f]\,.
\label{eq:rterm}
\eeq
This term has the effect of changing the normalization of the condensate
wavefunction.

We note that the kinetic equation (\ref{eq:kinetic}) for the thermal 
excitations is the same as that for gas of classical particles moving in the 
effective potential $U$. This is the basis of our numerical solution of 
this equation, which tracks the Hamiltonian dynamics of a swarm of test 
particles using a sympectic integrator.
Collisions are treated using a Monte Carlo evaluation of the collision
integrals. The GP equation can be propagated in time using a variety of 
methods, but we favour a Fast Fourier Transform split-step operator
technique. More details of the numerical methods can be found in Ref.\
\cite{jackson02b}. 

\section{Collective modes}

In this section we discuss the collective modes of the system, with particular
emphasis on the so-called transverse breathing mode in a cigar-shaped trap.
We begin by reviewing results for a pure condensate at temperature $T=0$. 
In this limit the condensate dynamics is described by (\ref{eq:GP-gen}) with
$\tilde{n}=0$ and $R=0$, i.e.,
\beq
  i\hbar \frac{\partial}{\partial t} \Phi(\brt) = \left (
 -\frac{\hbar^2 \nabla^2}{2m} + U_{\rm ext} (\br) + g n_c (\brt) \right) 
 \Phi(\brt)\,.
\label{eq:GP}
\eeq
The equilibrium solution can be found from (\ref{eq:GP}) by making the 
substitution $\Phi (\brt)=\Phi_0(\br) {\rm e}^{-i\mu t}$, where $\mu$ is the 
chemical potential. When the number of atoms is sufficiently large, the
kinetic energy term is dominated by the other terms and can be 
neglected. This defines
the Thomas-Fermi (TF) approximation and gives an equilibrium 
density profile of
\beq
 n_{c0} (\br) = \frac{1}{g} \left[ \mu - U_{\rm ext} (\br) \right].
\label{eq:TF-dens}
\eeq

The frequencies of condensate excitations can be found by 
linearizing (\ref{eq:GP}), and analytic solutions for the lowest modes
in the TF limit were first presented by Stringari \cite{stringari96}. 
In a spherical trap ($\lambda=1$), the modes can be labelled by the
angular momentum quantum numbers $(l,m)$. The lowest of these modes 
correspond to monopole $(0,0)$ and quadrupole $(2,0)$ excitations with 
frequencies $\sqrt{5} \omega_{\bot}$ and $\sqrt{2}\omega_{\bot}$,
respectively. In an anisotropic, axisymmetric trap these two 
modes are coupled to give the dispersion relation
\beq
 \omega_{\pm}^2 = \omega_{\bot}^2 \left( 2+\frac{3}{2}\lambda^2 \pm 
 \frac{1}{2}\sqrt{9\lambda^4-16\lambda^2+16} \right).
\label{eq:m0-modes}
\eeq
In Fig.~1 we plot these two solutions as a function of $\lambda$. The
two curves tend to the monopole and quadrupole frequencies of the
isotropic trap in the $\lambda\to 1$ limit. 
For the cigar-shaped condensate which will be of interest here ($\lambda
\ll 1$), the low frequency mode $\omega_{-}$ corresponds to a
breathing-like oscillation predominately along the axial ($z$) 
direction, while the high-lying mode $\omega_{+}$ is associated with 
the radial breathing mode. In the limit $\lambda \rightarrow 0$ the 
frequency of the latter approaches 
$2 \omega_{\bot}$. For example, in the experiment
of Ref.\ \cite{chevy02}, $\lambda \simeq 6.46 \times 10^{-2}$ and the 
transverse breathing mode has a frequency of $\omega_+ \simeq 2.00052 
\omega_{\bot}$.

These results are valid in the strongly interacting TF regime, but the 
non-interacting limit $g =0$ is also instructive. In this case the 
transverse quadrupole and breathing modes mentioned above both have a 
frequency of $2 \omega_{\bot}$. This clearly has interesting consequences 
for the breathing mode in the $\lambda=0$ limit, since it implies that the
frequency is the same in the non-interacting and strongly-interacting limits.
Indeed, calculations confirm \cite{pires00} that this is also true in the
intermediate regime, demonstrating the unusual property that the
frequency of this mode is independent of interactions. This behavior is reminiscent 
of that for a gas in a 2D harmonic trap interacting by means of a 
contact potential~\cite{kagan96,pitaevskii97}, where there exists an 
undamped breathing mode
with frequency $2 \omega_{\bot}$, independent of interaction strength, 
temperature, or statistics.

The thermal cloud dynamics above $T_c$ is given by (\ref{eq:kinetic}) 
with $n_c = 0$ and $C_{12}=0$, i.e.,
\beq      
 \frac{\partial}{\partial t} f (\bprt) + \frac{\bp}{m} \cdot \nabla
 f (\bprt) - \nabla U_{\rm ext} (\brt) \cdot \nabla_{\bp} f (\bprt)
 = C_{22} [f]\,,
\label{eq:boltz>T_c}
\eeq
where for simplicity we have also neglected the thermal cloud mean 
field term, $2g\tilde{n}$, since it is has a very small effect on the 
dynamics in the normal state. The lowest collective modes for the 
thermal gas can be conveniently found from (\ref{eq:boltz>T_c})
by taking moments. This procedure was used to study a classical gas in 
\cite{gueryodelin99}, though the results for a non-condensed Bose gas 
are essentially the same. 
The frequency and damping of the modes depend upon the mean collision rate
$\tau^{-1}$ relative to the frequency of the mode
$\omega$. In the purely collisionless regime $\omega \tau \rightarrow 
\infty$, and the frequencies of the transverse quadrupole and breathing modes
are both $\omega = 2 \omega_{\bot}$, while the damping is zero. In the
hydrodynamic limit ($\omega \tau \rightarrow 0$) the modes are also undamped,
but here the frequencies of the respective modes are $\omega_Q = \sqrt{2}
\omega_{\bot}$ and $\omega_M = \sqrt{10/3} \omega_{\bot}$ (where the 
latter is specifically for the cigar-shaped limit $\lambda 
\ll 1$)~\cite{gueryodelin99,griffin97}. In
general the modes will be damped through collisions, and will have a 
frequency intermediate between these two limits. In the experiment of
\cite{chevy02} the system resided in the near-collisionless regime, so
that both thermal cloud modes were only weakly damped with frequency
$\omega \simeq 2 \omega_{\bot}$. 

In order to understand the transverse breathing mode more fully it is
instructive to derive an equation of motion for the operator $R^2 =
\sum_i (x_i^2 +y_i^2)$. The expectation value of this operator for a
particular dynamical state gives the mean-squared radial size of the
cloud and it is therefore a suitable variable to describe the transverse
breathing mode of interest. For the Hamiltonian 
\beq
 H = \sum_i \frac{p_i^2}{2m} + \frac{m}{2} \sum_i \left[ \omega_{\bot}^2
 (x_i^2 + y_i^2) + \omega_z^2 z_i^2 \right] + \frac{g}{2} \sum_{i\neq j}
 \delta(\br_i - \br_j)\,,
\label{eq:hamil}
\eeq
we have the Heisenberg equation of motion
\beq
 \frac{{\rm d} R^2}{{\rm d} t} = \frac{i}{\hbar} [H,R^2] = \frac{Q}{m},
\label{eq:eqm1}
\eeq 
where the operator $Q$ is given by
\beq
Q=\sum_i (x_i p_{ix} + y_i p_{iy} + {\rm h.c.})\,.
\label{eq:Q}
\eeq
After differentiating (\ref{eq:eqm1}) again and evaluating the 
commutator $[H,Q]$ we obtain the result
\beq
 \frac{{\rm d}^2 R^2}{{\rm d}t^2} + (2\omega_{\bot})^2 R^2 = \frac{4}{m} 
 (H-T_z-U_z^{\rm ext}),
\label{eq:eqm2}
\eeq
where $T_z = \sum_i (p_{iz}^2/2m)$ is the kinetic energy associated with
motion in the $z$-direction and $U_z^{\rm ext} = 
\sum_i (m\omega_z^2 z_i^2/2)$. It is worth emphasizing that this
equation for $R^2$ is an operator identity which is equally valid
classically as it is quantum mechanically.

We now consider the expectation value of (\ref{eq:eqm2}) with respect to
some nonequilibrium density matrix. Denoting expectation values by
angular brackets, and noting that the energy, $E$, of the system is a 
constant of the motion, we have
\beq 
 \frac{{\rm d}^2 \chi}{{\rm d}t^2} + (2\omega_{\bot})^2 \chi = 
\frac{4}{m} [E-\la T_z \ra-\la U_z^{\rm ext} \ra]\,,
\label{eq:eqm3}   
\eeq
where we have defined $\chi \equiv \la R^2 \ra$.
For a strictly two-dimensional (2D) system, 
$\la T_z \ra =\la U_z^{\rm ext} \ra = 0$, and 
(\ref{eq:eqm3}) has the solution
\beq
 \chi = A \cos (2\omega_{\bot} t) + \bar\chi.
\label{eq:eqm3-sol}
\eeq 
In other words, $\chi$ oscillates about $\bar\chi=E/(m\omega_{\bot}^2)$ 
with frequency $2\omega_{\bot}$. 
This behaviour was first noted by Pitaevskii and Rosch in their analysis
of the 2D system.
For three-dimensional systems, $\la T_z \ra$ and $\la U_z^{\rm ext} \ra$
are themselves dynamical variables. Their appearance in (\ref{eq:eqm3})
will in general lead to a shift of the frequency of the transverse 
breathing mode from $2\omega_\perp$, except in certain limiting cases.

The equation of motion for mean values in (\ref{eq:eqm3}) is also valid
when the dynamics is governed by the time-dependent GP equation. However
in this case, the average of an operator ${\cal O}$ is defined as $\la
{\cal O} \ra = \int {\rm d}\br\, \Phi^*(\br,t) {\cal O} \Phi(\brt)$.
In the limit of a cylindrical trap,
$\lambda\rightarrow 0$ and $\la U_z^{\rm ext}\ra = 0$. If we then
consider an axially symmetric mode which is spatially invariant along 
the length of the trap, we have $\la T_z\ra =0$ and (\ref{eq:eqm3})
implies that a mode at the frequency $2\omega_\perp$ exists. This
particular mode was previously identified as the transverse breathing 
mode in a cylindrical trap. It should be emphasized that this conclusion
is based on the approximate GP description. In the many-body treatment,
 $\la T_z\ra
\ne 0$ and the existence of a $2\omega_\perp$ mode then depends on
$\la T_z\ra$ being constant. It is not known whether a dynamical state 
with this property exists in general.

In applying (\ref{eq:eqm3}) to a semiclassical gas we interpret
the expectation value of a physical observable ${\mathcal{O}}$ as the 
phase space
average $\la{\mathcal{O}}\ra=\int ({\rm d}\bp{\rm d}\br/h^3) 
{\mathcal{O}(\bp,\br) } f(\bprt)$. Two limits are of interest: the
collisionless and the hydrodynamic. In the former case one can 
use (\ref{eq:boltz>T_c}) to show that
${\rm d} (\la T_z \ra + \la U_z^{\rm ext}\ra)/{\rm d} t = 0$ 
for a separable confining potential. Eq. (\ref{eq:eqm3}) then implies
that a collisionless gas has a transverse breathing mode at the 
frequency $2\omega_{\bot}$, as noted earlier.

Once collisions are included, $(\la T_z \ra + \la U_z^{\rm ext}\ra)$ is
no longer a constant of the motion and the frequency of the transverse
breathing mode will change. We can illustrate this for a cylindrical
trap ($\omega_z =0$) in the hydrodynamic limit.
In this regime, collisions are sufficiently frequent to ensure
local thermodynamic equilibrium, so that the distribution function is of
the Bose form
\beq
 f(\bprt) = \frac{1}{{\rm e}^{\beta(\brt) \{ [\bp-m\bv(\brt)]^2/2m 
 - \mu(\brt) \} - 1}}\,,
\label{eq:leqmbose}
\eeq
where $\beta(\brt)$, $\mu(\brt)$ and $\bv(\brt)$ are all local
hydrodynamic variables. For the transverse breathing mode the
local velocity field has the form $\bv(\brt) = a\br_\bot$, where
$\br_\bot$ is a
vector in the radial direction, tranverse to the axis of the trap.
Using (\ref{eq:leqmbose}) one can show that $\la T_z \ra = \la T \ra / 3$ 
and $E = \la T \ra + m\omega_{\bot}^2 \chi/2$, where we have neglected 
the nonlinear contribution of the velocity field to the kinetic 
energy $\la T \ra = \la p^2/2m \ra$ of the mode. By substituting these 
expressions into 
(\ref{eq:eqm3}) one finds that the transverse breathing mode has
frequency $\omega_M = \sqrt{10/3} \omega_{\bot}$, in agreement with 
other calculations in this limit \cite{gueryodelin99,griffin97}. This
provides a concrete example of a situation in which a non-constant
$\la T_z \ra$ leads to a shift in the mode frequency from 
$2\omega_\perp$. It is therefore clear that
a mode at this frequency is not a universal property of a trapped gas
in the cylindrical geometry.

\section{Numerical results}

The behavior of the condensate at $0<T<T_c$ is influenced by the
presence of thermal excitations. Normally this leads to a decay of 
condensate oscillations, which in the near-collisionless regime arises 
through Landau damping. There is also an associated frequency shift.
However, in the experiment of \cite{chevy02}, the transverse breathing 
mode was found to be very weakly damped, with a frequency almost 
independent of temperature. This is in stark contrast to the much larger
damping rate found in their measurements of the quadrupole mode, as well
as in other experiments~\cite{jin97,stamper-kurn98,marago01}.
In this section we present results of both the breathing and quadrupole
modes for the system studied in Ref.~\cite{chevy02}. We show that the 
observed behavior of the breathing mode is a consequence of the
degeneracy of the condensate and thermal cloud mode frequencies coupled
with the fact that both components are initially excited. 

The experimental scheme used to excite the system was to suddenly 
change the radial trap frequency and then to reset it to its 
original value after some short time $\tau$. This stepped excitation 
can be represented by
\beq
 \omega_{\bot}' (t) = \omega \{1+\alpha [\Theta (t) - \Theta (t-\tau)] \},
\label{eq:excite}
\eeq
where we take $\omega_{\bot} = 2\pi \times 182.6\,{\rm Hz}$, $\alpha = 
0.26$, and $\omega_{\bot}\tau = 0.172$. Although this is a relatively
simple method for exciting condensate oscillations, in finite 
temperature studies it has the disadvantage of also exciting the 
thermal cloud. This aspect is relatively unimportant when the 
oscillation frequencies of the condensate and thermal cloud are 
sufficiently different, but when they are similar it can have a profound
effect on the behavior of the system. An important implication is that
the usual perturbative calculation of Landau damping which assumes the 
thermal cloud to be at 
rest~\cite{morgan00,rusch00,giorgini00,guilleumas00,reidl00} are no
longer applicable. 

To illustrate this point, we show in Fig.\ \ref{m0plot}(a) the result 
of a simulation at $T=125\,{\rm nK}$ (as compared to the experimentally
measured
critical temperature of $T_c \simeq 290\,{\rm nK}$). We plot the radial
moments for the condensate, $\chi_c$, and thermal cloud, $\chi_n$, as a 
function of time, where each has been divided by its initial value to 
give the relative oscillation amplitude. One sees that both components 
respond to the stepped excitation by oscillating in phase with 
approximately equal amplitudes. In addition, the oscillation frequency
is very close to $2\omega_{\bot}$ and the damping rate is very small.
As discussed earlier, the frequency is that expected for the condensate
for $\lambda \simeq 0$ and for the thermal cloud in the collisionless
regime. The common in-phase oscillation of the two components is further
reinforced by mean-field interactions. The lack of relative motion
between the two components accounts for the fact that Landau damping is
not effective. Evidence supporting this interpretation
is provided in Fig.\ \ref{m0plot}(b), where we plot the
result of exciting only the condensate by imposing a radial 
velocity field, $\bv_c \propto x \hat{\bf i}+y\hat{\bf j}$.
In this case the condensate oscillates initially in the presence 
of a {\it stationary} thermal cloud and as a result, Landau
damping is fully active, giving rise to the much larger damping rate
seen here. 

To quantify the frequency and damping rate, we fit the condensate 
oscillation to an exponentially decaying sinusoid 
$A \cos (\omega t + \varphi) {\rm e}^{-\Gamma t} + C$. 
Fig.\ \ref{freqdamp} compares our results 
at different temperatures with the experimental data. We see very good 
agreement between simulations and experiment for both the frequency and 
damping rate. As discussed in \cite{jackson02c}, the damping that occurs
is essentially due to $C_{12}$ and $C_{22}$ collisional processes. This
is unlike the situation for other modes in the near-collisionless regime
\cite{jackson02b,jackson01} where in fact Landau damping is the 
dominant damping mechanism. As indicated in Fig.\ \ref{m0plot}(b), 
Landau damping of the breathing mode only appears when the two 
components are moving relative to each other.

Recently, results from a calculation by Guilleumas and Pitaevskii 
appeared
\cite{guilleumas02} that point to a different conclusion. They performed
a perturbation theory calculation similar to \cite{guilleumas00}, but
for a cylindrically trapped condensate. In this geometry the excitation
spectrum, as obtained from the Bogoliubov equations, is characterized
by a series of curves continuous in $k$, the axial wavevector, with
each curve corresponding to distinct radial ($n$) and azimuthal angular
momentum ($m$) quantum numbers \cite{fedichev01}. Calculation of the 
Landau damping rate then consists of evaluating the following expression
\beq
 \Gamma = \frac{\pi}{\hbar^2} \sum_{ij} |A_{ij}|^2 (f_i - f_j)
 \delta (\omega_j - \omega_i - \omega_{osc})\,,
\label{eq:pert-damp}
\eeq   
where the delta function reflects energy conservation of the decay 
process, in which a quantum of oscillation $\hbar \omega_{\rm osc}$
is annihilated, and the $i$-th excitation transformed into the 
$j$-th. The probability of this transition is determined by 
the matrix element $A_{ij}$ (as defined in Ref.\ \cite{guilleumas02}),
and the distribution function $f$. The thermal cloud is assumed 
to be in thermodynamic equilibrium, so that the states $i$, $j$
are thermally occupied according to the Bose distribution
$f_i=[\exp(\hbar\omega_i/k_B T)-1]^{-1}$. Using this method they 
calculated damping rates for the lowest $m=0$ and $m=2$ modes in the
$k=0$ limit, which corresponds to the transverse modes in the elongated
3D geometry. We shall discuss the $m=2$ results later, but it is 
interesting that they found a damping rate for the $m=0$ which is an 
order of magnitude {\em smaller} than the experimentally measured value.
Since this calculation assumes a stationary thermal cloud, it is in
fact {\em two} orders of magnitude smaller than our result for the 
analogous situation where only the condensate is excited initially
[Fig.\ \ref{m0plot}(b)]. 

The discrepancy between the two results could arise for a variety
of reasons. It may be that the semiclassical approximation, as
employed in our calculations, is not valid for this particular case. The
small damping rate in Ref.\ \cite{guilleumas02} is partly the result of
the small number of transitions that satisfy energy conservation. This
constraint is less severe in the semiclassical approximation in which
the excitations are continuous in {\it both} the axial and radial
directions. The damping rate seen in [Fig.\ \ref{m0plot}(b)] might
therefore be an overestimate.  On the other hand, there is 
also the possibility that the small damping rate is an artifact of
the calculational method employed for the infinite cylinder. In the 
spherical trap discussed by the same authors \cite{guilleumas00},
energy can never be conserved precisely in a 
transition due to the discreteness of the energy levels. So in practice
the Landau damping rate (\ref{eq:pert-damp}) is calculated by replacing
the delta functions with Lorentzians of finite width. In a previous 
publication \cite{jackson02b} we found good agreement
between this approach and our simulations. This procedure would have to
be employed for any 3D system with a finite value of $\omega_z$. 
Since the excitation spectrum is a continuous function of $k$
for an infinite cylinder, this procedure for dealing with discrete
excitations is not required. The neglect of off-energy shell
transitions in this case may be leading to an underestimate of the
damping rate as compared to the finite trap situation.
In order to address these questions it would be useful
to extend the perturbative calculations of \cite{guilleumas00} 
to elongated condensates. One could then study the variation of the
damping with anisotropy to see whether the results begin to diverge 
from the semiclassical simulations at some point. We are currently 
exploring this possibility.

As we discussed above, the small damping rates and frequency shifts 
shown in Fig.\ \ref{freqdamp} are a consequence of the fact that there
is no relative motion of the condensate and thermal cloud in the
transverse breathing mode. This behaviour does not occur if the 
two components tend to oscillate at different frequencies as is the
case for the $m=2$ transverse quadrupole mode. To demonstrate this
we have performed simulations for the $m=2$ mode by imposing identical
velocity fields of the form $\bv_{\{c,n \}} \propto x \hat{\bf i}-
y\hat{\bf j}$ on the two components. An example of the subsequent 
oscillations of the condensate and thermal cloud is shown in 
Fig.\ \ref{m2plot}. One sees that the frequencies are
indeed quite different in this case -- the condensate oscillates with a 
frequency approximately equal to  $\sqrt{2} \omega_{\bot}$, while the 
thermal cloud oscillates at a frequency close to $2 \omega_{\bot}$. 
As for the $m=0$ 
mode, we fit the condensate data to a damped sinusoid, and plot the 
frequency and damping rate as a function of temperature in Fig.\ 
\ref{freqdamp2}. In contrast to the $m=0$ data in Fig.\ \ref{freqdamp},
the $m=2$ mode shows a much larger shift in frequency, as well 
as an order of magnitude increase in damping rate. The relative motion 
between the condensate and thermal cloud in this case accounts for these
increases. Indeed, from the perspective of the condensate, the thermal 
cloud motion averages over the course of the simulation to that of a 
stationary cloud. This is confirmed by simulations where the thermal 
cloud is not excited initially. As seen in Fig.\ \ref{freqdamp2}, 
these simulations exhibit very similar results to when
both components are perturbed, especially when $T \leq 150\,{\rm nK}$. 
The differences between the two schemes at higher temperatures mainly
reflect statistical variations of the results from one simulation to
the next.

Interestingly, when Guilleumas and Pitaevskii \cite{guilleumas02} 
investigated the
$m=2$ mode for the cylindrical limit, they found damping rates
in line with our results in Fig.\ \ref{freqdamp2}, especially at higher 
temperatures.  There are some 
discrepancies at low temperatures, but it must be remembered that 
precise agreement would not be expected between the two calculations
due to differences in geometry, parameters and calculational method. 
Nonetheless, it
is intriguing that such good agreement should be found here and not for
the $m=0$ mode. Again, further study is needed. Experiments for the 
$m=2$ mode would also be of interest, since comparison to the results
in Fig.\ \ref{freqdamp2} would provide further information regarding 
the reliability of our theory.

Finally, we discuss another method for breaking the degeneracy of the
condensate and thermal cloud modes.
As seen in Fig.\ \ref{lambdaf}, increasing the parameter 
$\lambda$ leads to a rise in the breathing mode frequency $\omega_+$,
while at the same time the thermal cloud frequency remains near 
$2\omega_{\bot}$. Thus,
as the trap becomes more isotropic the condensate frequency moves away
from degeneracy, and one would expect the damping of the breathing 
mode to correspondingly increase. To check this
we have performed simulations for larger values of $\lambda$. An
example of the response to a stepped excitation (\ref{eq:excite}) is 
shown in Fig.\ \ref{anisospec} for $\lambda =0.75$; the geometric mean
of  the trap frequency is chosen to be the same as in the ENS experiment
($\bar{\omega}=\lambda^{1/3} \omega_{\bot} =2\pi\times 73.3\,{\rm Hz}$).
Fig.~\ref{anisospec}(b) shows the time dependence of the thermal cloud
moment $\chi_n$. It exhibits a damped oscillation at essentially one
frequency; a single-mode fit to the data yielded a frequency of
$\omega/\omega_\bot \simeq 2.007$ and a damping rate of
$\Gamma/\omega_\bot \simeq 0.019$. The latter is approximately four
times larger than found for the ENS geometry. The behavior of $\chi_c$ 
in Fig.~\ref{anisospec}(a) is more complex due to the fact that both of
the $\omega_\pm$ modes are being excited. In fact, a combination of two
damped sinusoids provided a very good fit to the time dependence shown
and yielded the following fit parameters: $\omega_-/\omega_\bot \simeq
1.12$, $\Gamma_-/\omega_\bot \simeq 0.037$; $\omega_+/\omega_\bot \simeq
2.03$, $\Gamma_+/\omega_\bot \simeq 0.024$. The frequencies are quite
close to the values given by (\ref{eq:m0-modes}) for $\lambda = 0.75$.
More importantly, we see that the relative damping rate of the
$\omega_+$ mode is approximately an order of magnitude larger than that
of the transverse breathing mode from which it evolves. Thus even a
relatively small difference in the frequencies of the condensate and
thermal cloud modes is sufficient to significantly enhance the Landau
damping rate. This fact emphasizes the very special nature of the
transverse breathing mode in the $\lambda \to 0$ limit.

\section{Summary}

In this paper we have investigated the transverse oscillations of a 
Bose gas in the context of semiclassical finite temperature theory.
Analytical results were derived in the cylindrical limit for the dynamics
of the condensate at $T=0$ and thermal cloud for $T>T_c$. Simulations are 
used to study the dynamics in an elongated trap at $0<T<T_c$, for the 
parameters of the experiment of \cite{chevy02}. Our results for 
the frequency and damping of the $m=0$ mode demonstrate very good agreement 
with this experiment. However, comparison to the results of a related 
theoretical study in \cite{guilleumas02} raises several interesting 
questions. We also study the $m=2$ mode, and find a much larger damping
rate and frequency shift associated with the presence in this case of 
relative motion between the condensate and thermal cloud. Finally, we study
the effect on the $m=0$ mode of decreasing the trap anisotropy, and find
that the damping increases dramatically as soon as
the accidental degeneracy between the condensate and thermal cloud 
frequencies is lifted.

\acknowledgements
We acknowledge financial support from the Natural Sciences and
Engineering Research Council of Canada, and
the use of the HPCVL computing facilities at Queen's. We thank L.
Pitaevskii for valuable discussions.

\begin{figure}
\centering
\psfig{file=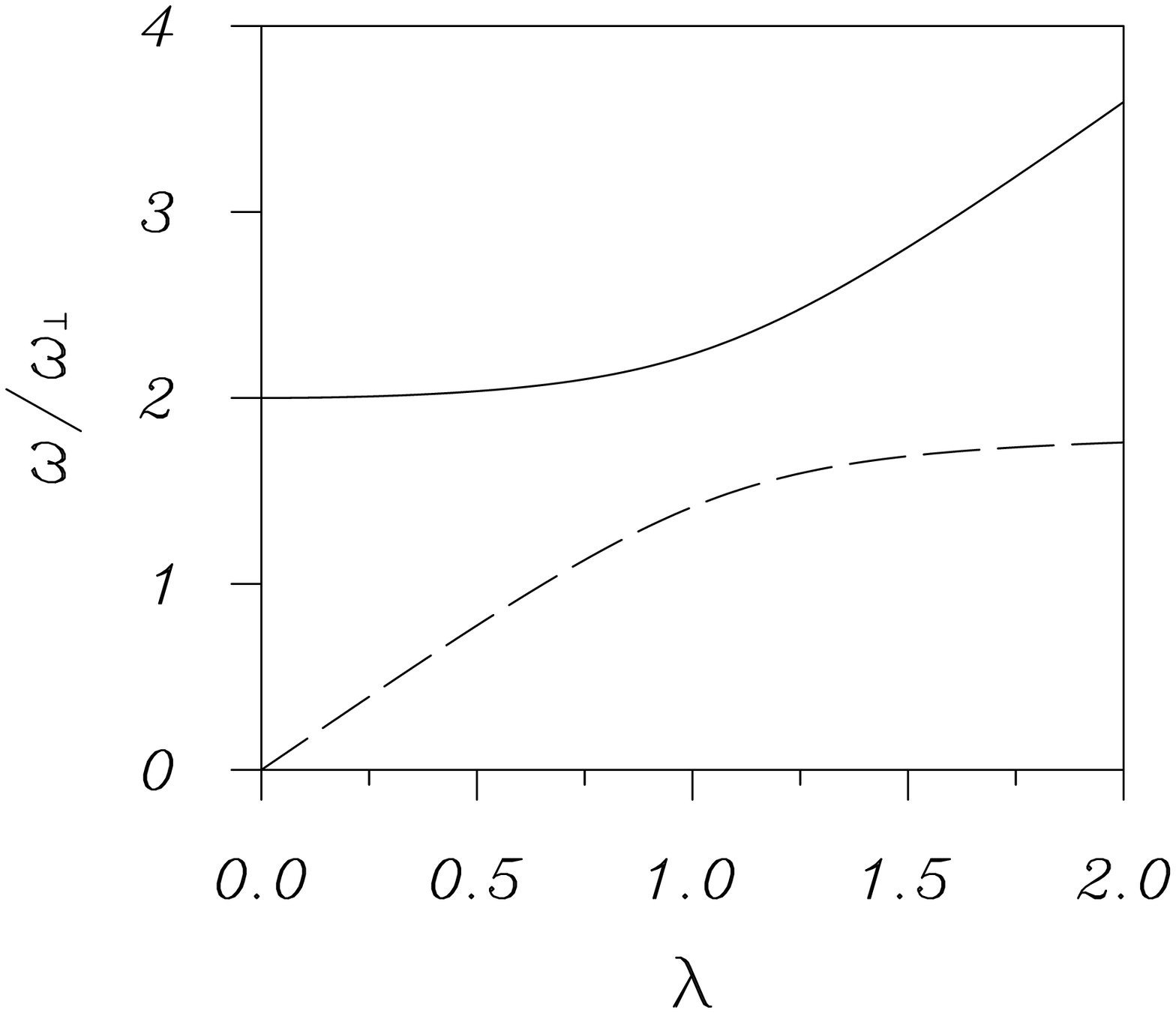, scale=0.8, bbllx=70, bblly=160, 
 bburx=575, bbury=610}
\caption{\label{lambdaf} 
 Frequencies of the lowest $m=0$ condensate modes in the Thomas-Fermi 
 limit, given by (\ref{eq:m0-modes}), as a function of the anisotropy 
 parameter $\lambda=\omega_z/\omega_{\bot}$. The higher
 mode $\omega_+$ (which corresponds to the transverse breathing mode in
 the $\lambda \rightarrow 0$ limit) is labelled by the solid line, 
 while $\omega_-$ is denoted with the dashed line.} 
\end{figure}

\begin{figure}
\centering
\psfig{file=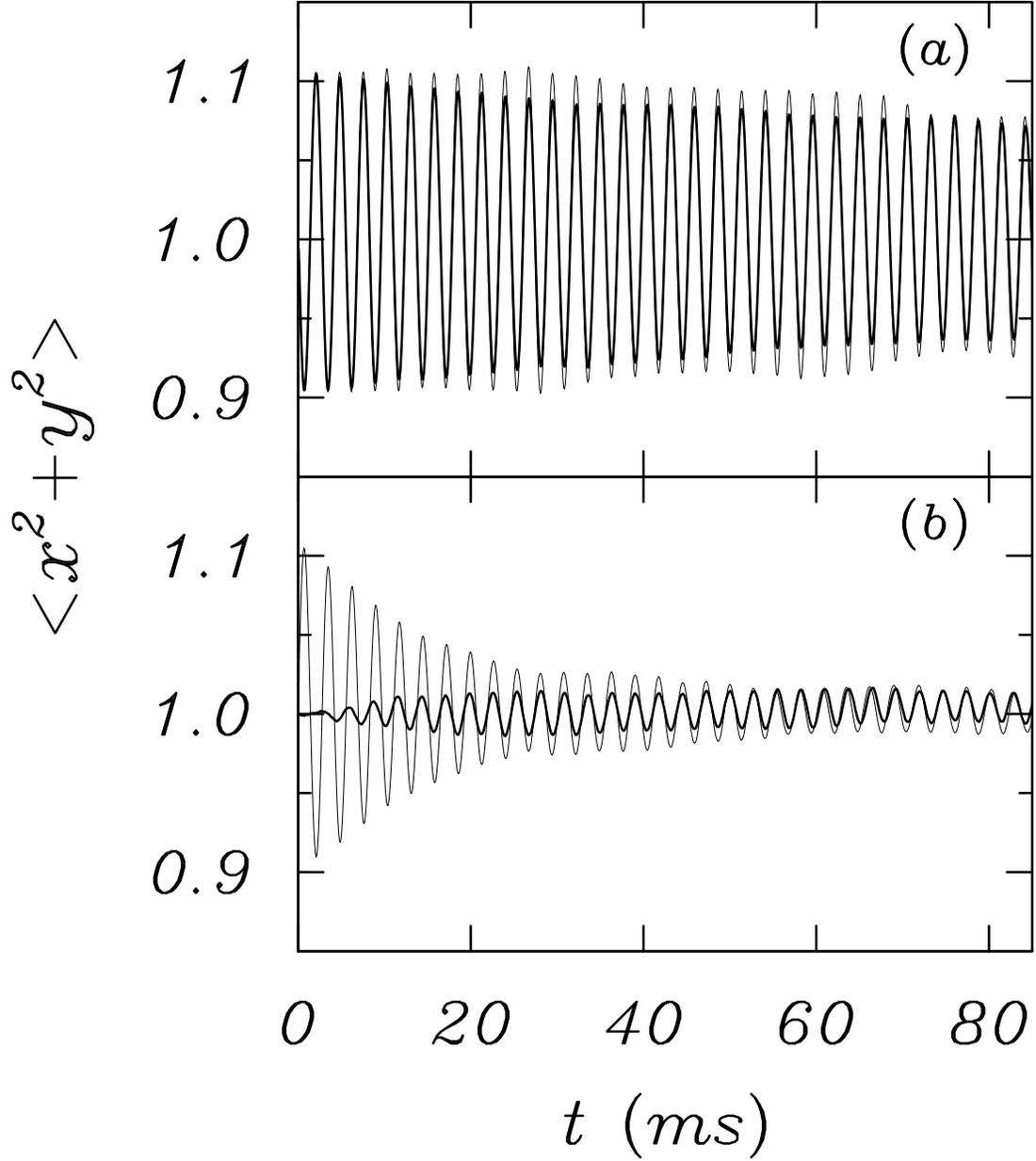, scale=0.8, bbllx=40, bblly=75, 
 bburx=560, bbury=680} 
\caption{\label{m0plot} 
 Time-dependent radial moments for the condensate,
 $\chi_c$ (narrow line) and thermal cloud, $\chi_n$ (bold line), divided
 by the corresponding values at $t=0$. The figures are for a temperature
 of $T=125\,{\rm nK}$, and show the result of (a) exciting the system
 using the ``tophat'' perturbation scheme employed experimentally 
 (\ref{eq:excite}), and (b) exciting the condensate only by initially 
 imposing a velocity field of the form $\bv_c \propto x \hat{\bf i} + 
 y \hat{\bf j}$.}
\end{figure} 

\begin{figure}
\centering
\psfig{file=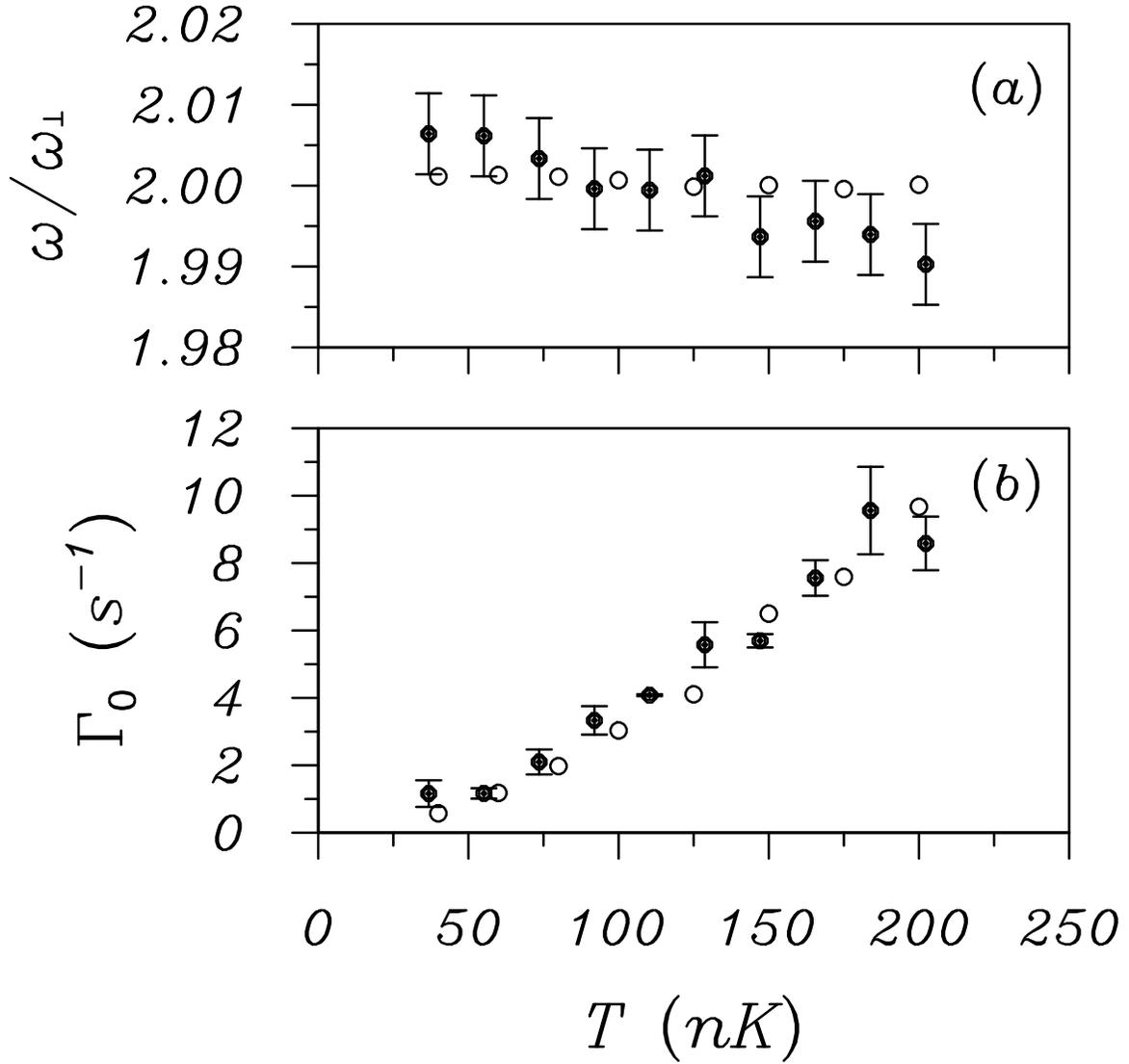, scale=0.8, bbllx=25, bblly=100, 
 bburx=580, bbury=650}
\caption{\label{freqdamp} Frequency (a) and damping rate (b) of the 
 condensate transverse breathing mode. Our results (open circles) are 
 compared to the 
 experimental data of {\protect \cite{chevy02}} (solid circles), where the
 simulation parameters and excitation scheme are chosen to match the 
 conditions of the experiment.} 
\end{figure}

\begin{figure}
\centering 
\psfig{file=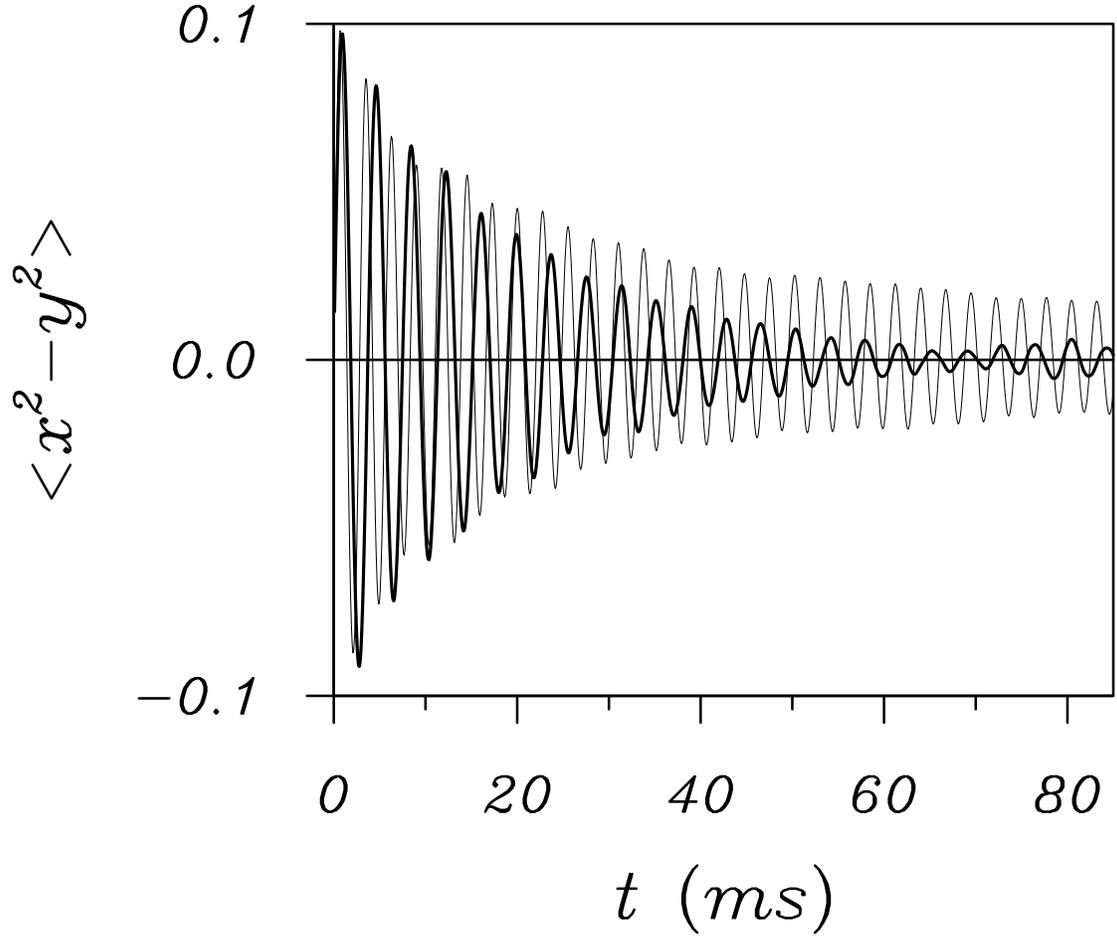, scale=0.8, bbllx=25, bblly=145, 
 bburx=555, bbury=610}
\caption{\label{m2plot} Time-dependent quadrupole moments 
 $\la x^2-y^2 \ra$ for the condensate (bold line) and thermal cloud 
 (narrow line), where the moments have been scaled so that they overlay one
 another. The plots are for a temperature of 
 $T=125\,{\rm nK}$, and show the result of initially imposing a velocity
 field of the form  $\bv_{\{c,n\}} \propto x \hat{\bf i}-y \hat{\bf j}$
 on both components.}
\end{figure}

\begin{figure}
\centering
\psfig{file=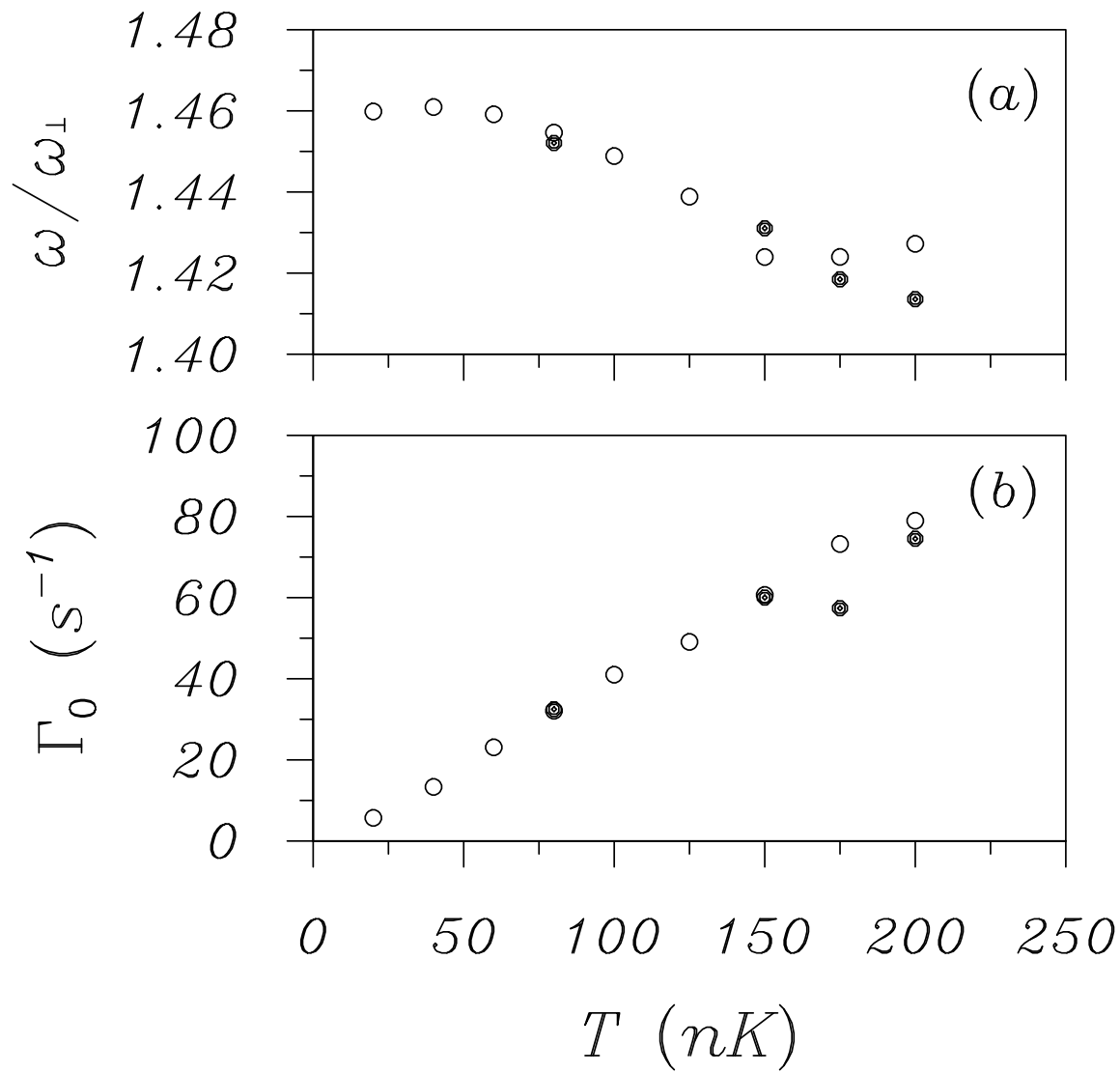, scale=0.8, bbllx=25, bblly=100, 
 bburx=580, bbury=650}
\caption{\label{freqdamp2} Frequency (a) and damping rate (b) of the 
 condensate transverse quadrupole mode, where we show the results of 
 initially exciting both components (open circles) and the condensate only
 (bullets) by imposing a velocity field. Simulation parameters are the 
 same as in Fig.\ \ref{freqdamp}.} 
\end{figure}

\begin{figure}
\centering
\psfig{file=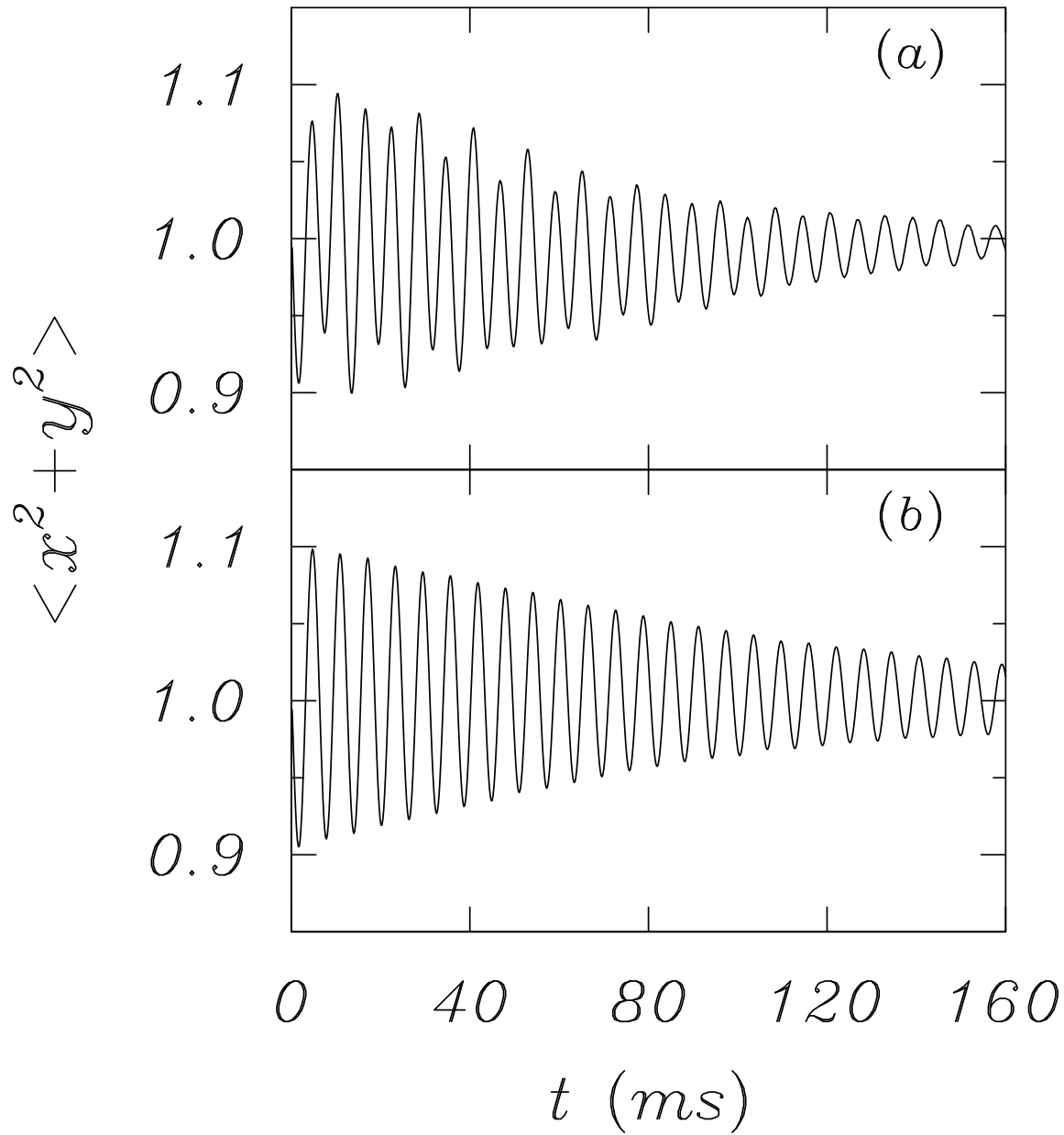, scale=0.8, bbllx=115, bblly=80, 
 bburx=560, bbury=680}
\caption{\label{anisospec} 
 Time-dependent radial moments for (a) the condensate and (b)
 the thermal cloud, divided by the corresponding values at $t=0$,
 for a less anisotropic trap. 
 The figures are for a temperature of $T=125\,{\rm nK}$, and show the 
 result of perturbing the system using the stepped excitation in 
 (\ref{eq:excite}). The trap parameters are $\lambda = 0.75$ and
 $\bar \omega = 2\pi \times 73.3$ Hz.}
\end{figure}

\end{document}